\documentclass[12pt,showpacs,showkeys,preprintnumbers,nofootinbib]{revtex4-1}

\usepackage{amssymb}
\usepackage{verbatim} 
\usepackage{subfig}
\usepackage{graphicx}
\usepackage{amsthm}
\usepackage{color}
\usepackage{soul}

\begin{document}

\title{Impact of network randomness on multiple opinion dynamics}

\author{Vivian Dornelas $^{1}$, Marlon Ramos $^{2}$ and Celia Anteneodo $^{1}$}

\address{
$^{1}$ \quad Department of Physics, PUC-Rio, Rio de Janeiro, Brazil; viviandnunes@gmail.com; celia.fis@puc-rio.br\\
$^{2}$ \quad Gleb Wataghin Institute of Physics, Universidade Estadual de Campinas, S\~ao Paulo, Brazil; marlonf.ramos@gmail.com}

\begin{abstract}

People often face the challenge of choosing among different options with similar attractiveness.
To study the distribution of preferences that emerge in such situations, a useful approach is to simulate opinion dynamics on top of complex networks, composed by nodes (individuals) and their connections (edges), where the state of each node can be one amongst several opinions including the undecided state.
We use two different dynamics rules: the one proposed by Travieso-Fontoura (TF) and the plurality rule (PR), which are paradigmatic of outflow and inflow dynamics, respectively.
We are specially interested in the impact of the network randomness on the final distribution 
of opinions. For that purpose, we consider Watts-Strogatz networks, which possess the small-world property, and where randomness is controlled by a probability $p$ of adding random shortcuts to an initially regular network.
Depending on the value of $p$, the average connectivity $\langle k \rangle$, and the initial conditions, the final distribution can be basically (i) consensus, (ii) coexistence of different options, or (iii) predominance of indecision.
We find that, in both dynamics, the predominance of a winning opinion is favored by increasing the number of reconnections (shortcuts), promoting consensus. In contrast to the TF case, in the PR dynamics, a fraction of undecided nodes can persist in the final state. In such cases, a maximum number of undecided nodes occurs within the small-world region, due to ties in the decision group.

\end{abstract}

%
%
%
%


\maketitle

\section{Introduction}
\label{sec:intro}
  
Most opinion models proposed in the sociophysics literature~\cite{castellano,galam2012sociophysics,sen2013sociophysics,galam2008} 
consider a binary variable, since many problems can be analyzed through 
the assumption of two single choices (e.g., for and against).
However, in many everyday situations, we have to choose an option among several 
available ones with similar attractiveness, for example, 
choosing a movie, restaurant or buying a simple product in a supermarket.
When we face such situations, without clear knowledge of the products offered, 
we tend to be influenced by friends,  family and other contacts. 
There may be cases where each contact suggests a different product, 
and we remain undecided. 
Despite these are common situations, few studies of social dynamics 
address the possibility of multiple choices~\cite{Chen2005a,Holme2006a,axelrod1997,vazquez2007non,calvao2016role}. 
Therefore, there are still many open questions, one of them is about the effect of contact network topology and, 
particularly, its level of randomness. 
This scenario motivates the present work. 

In a network, sites represent individuals and edges the possibility of interaction between the linked sites. 
To each site one attributes a state, that can evolve through the interaction with contact neighbors. 
As a prototypical network of connections between individuals, we use the network 
proposed by Watts and Strogatz (WS)~\cite{watts1998collective} 
because it produces the small-world (SW) property that is observed in 
many real social networks. 
In this network, it is possible to adjust the level of randomness, 
through a parameter $p$, 
relinking connections starting from a regular lattice.

\begin{figure}[!ht]
\centering
\subfloat[TF dynamics]{
\includegraphics[width=0.2\textwidth]{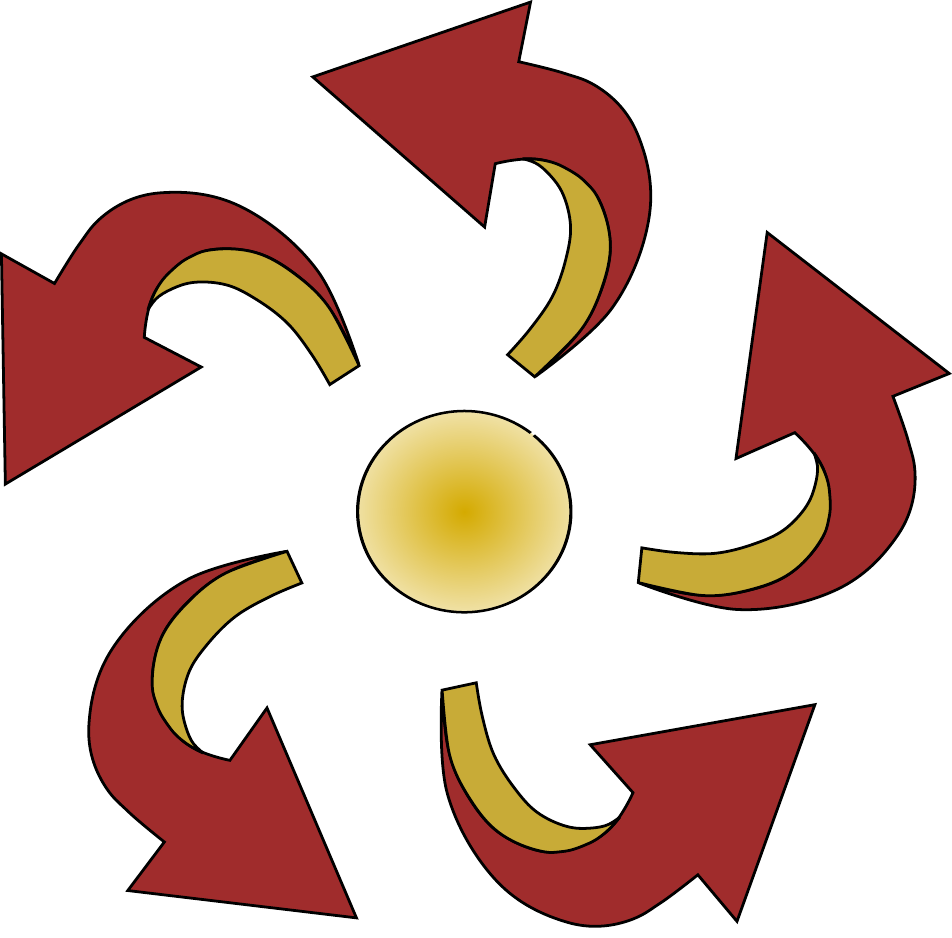}
}
\qquad
\subfloat[PR dynamics]{
\includegraphics[width=0.2\textwidth]{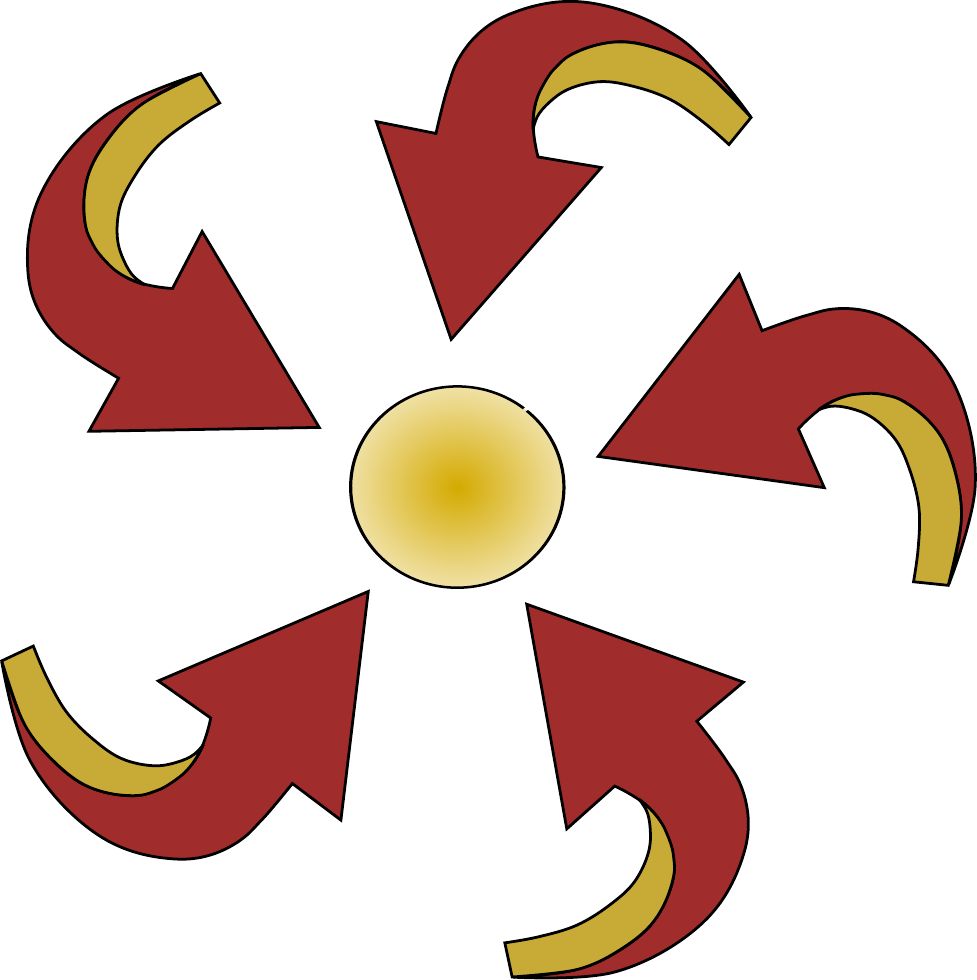}
}
\caption{
	Pictorial representation of the opinion contagion rules, where the propagation occurs  
	(\textbf{a})  from the individual to his/her neighbors (TF rule) or   
	(\textbf{b}) from the local neighborhood to a given individual (PR rule).
	}
\label{fig:flows}
\end{figure}

We will consider that the changes of opinion are governed by  rules appropriate 
to our problem of interest. 
Then, we  consider variants of two paradigmatic rules of opinion dynamics, 
both contemplating the possibility of multiple choices, 
as well as the undecided state. 
One of the rules is a proposal by Travieso and Fontoura~\cite{travieso2006spread} (TF), 
where the ``contagion'' of preferences occurs from an individual 
towards his/her neighbors in the contact network (outflow dynamics). 
The other  one is a plurality rule~\cite{calvao2016role} (PR), where the transmission 
of preferences occurs in the opposite direction, from the neighborhood towards the individual (inflow dynamics).
Figure~\ref{fig:flows} presents a pictorial representation of both rules. 
Their precise definitions will be given in Sec.~\ref{sec:dynamics}. 
Moreover, for TF case, the update is done asynchronously, 
but for  PR case, two forms of update, asynchronous and synchronous, are considered.

We will see that both rules can give rise to different final configurations, 
such as coexistence of many preferences, consensus, or yet, cases in which the quantity  
of undecided individuals is expressive.
The final distribution of opinions in the population will be characterized basically 
by  the $f_w$ fraction  of individuals who have adopted the alternative 
with more adepts and by the $f_0$ fraction of undecided individuals. 
These two quantities can be influenced by the randomness $p$ of the network 
and by its average connectivity $\langle k \rangle$, or even, 
by the initial conditions. Therefore, we will vary these factors to show their impact 
on the final distribution of opinions.

The paper is organized as follows. 
The networks used and the dynamic rules are defined in Secs.~\ref{sec:networks} and 
Sec.~\ref{sec:dynamics}, respectively. The results of our analysis are presented in 
Sec.~\ref{sec:results} and final remarks are done in Sec.~\ref{sec:conclusions}.

\section{Watts-Strogatz  networks}
\label{sec:networks}

To create a Watts-Strogatz (WS) network, we follow the standard procedure~\cite{watts1998collective},   
starting from a regular ring of $N$ nodes, each one with connectivity $k$, 
and using a rewiring probability $p$ ($0\leq p\leq 1$).

\begin{figure}[!ht]
	\includegraphics[width=0.45\textwidth]{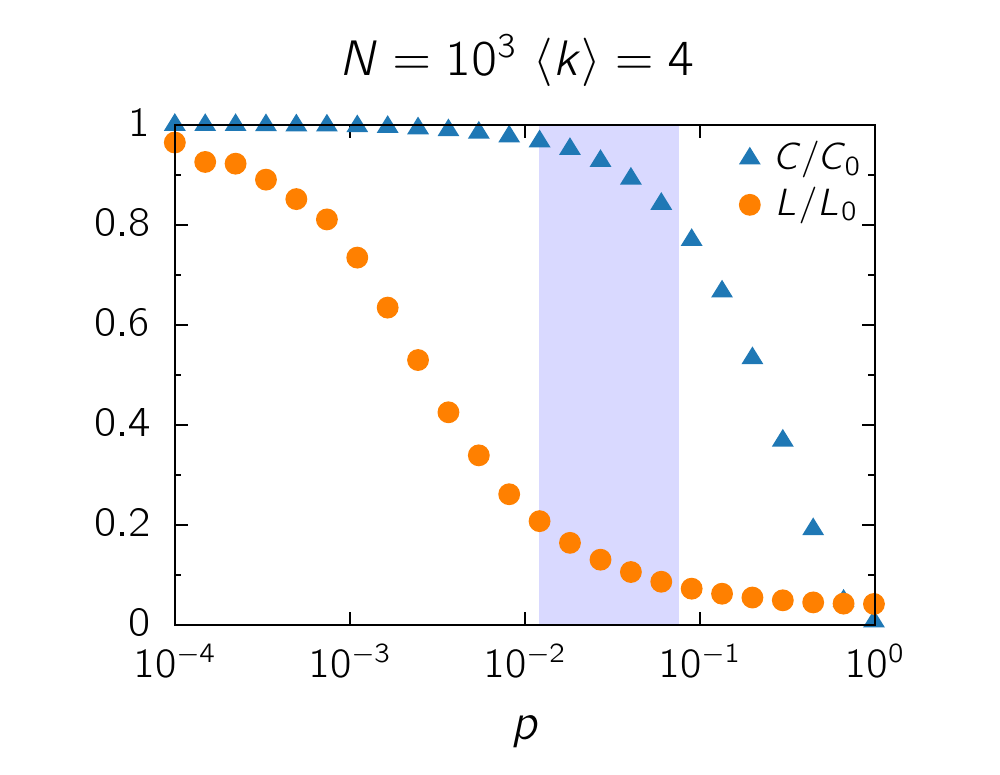}\hspace{-0.7cm}
	\includegraphics[width=0.45\textwidth]{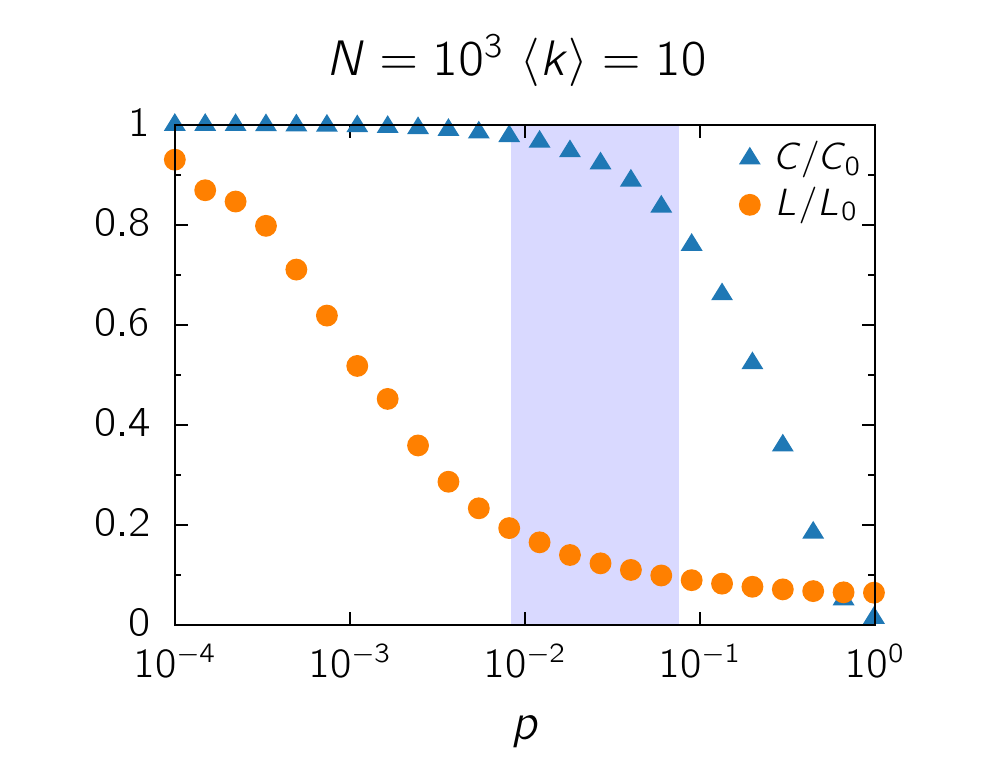}

	\includegraphics[width=0.45\textwidth]{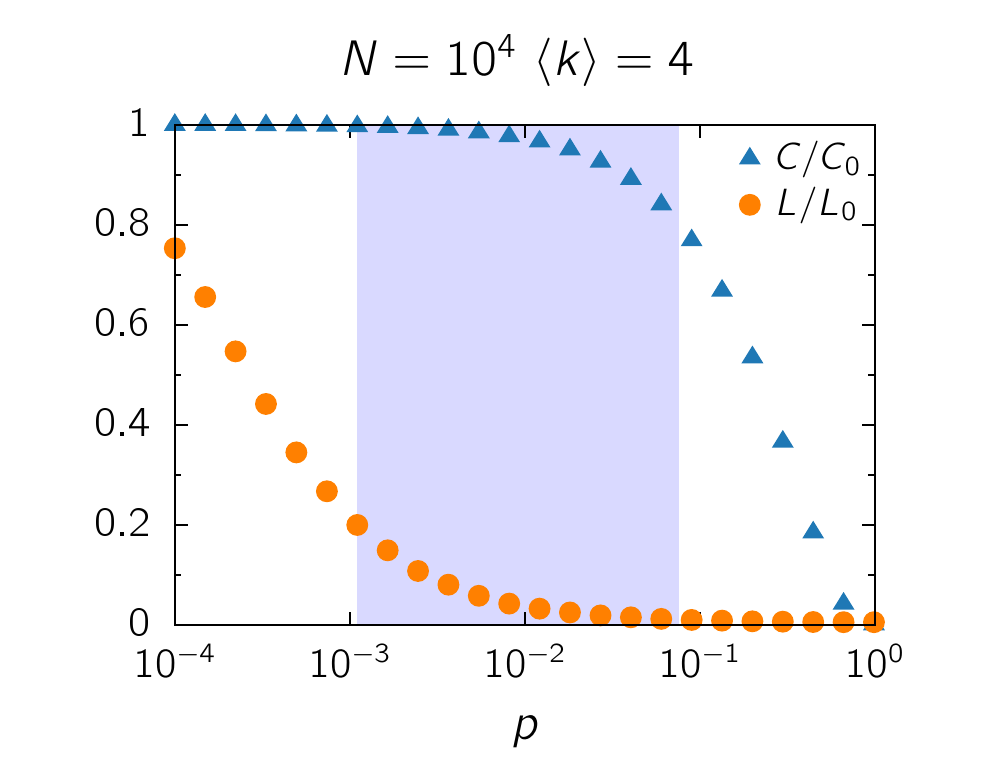}\hspace{-0.7cm}
	\includegraphics[width=0.45\textwidth]{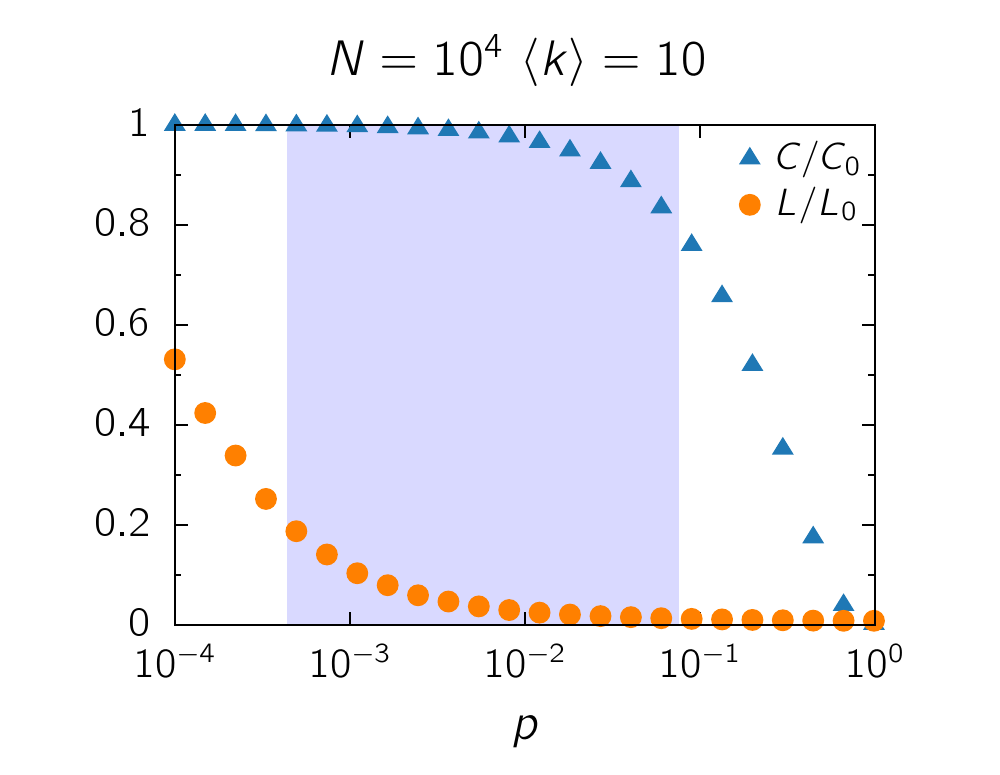}
	\caption{ $C/C_0$ and $L/L_0$ as a function of the randomness of the network $p$, 
	where $C_0$ and $L_0$ are, respectively, the agglomeration coefficient and 
	the smallest mean path when $p=0$. 
	Each symbol represents the average over 100 samples, 
	for networks with different values of $N$ and $ \langle k \rangle$, indicated in each panel.
The shadowed region is delimited by vertical lines given by  $p_1(\epsilon)$ and $p_2(\epsilon)$, 
according to Eq.~(\ref{eq:p1p2}).}
	\label{fig:CeL}
\end{figure}

Two useful measures of a network structure are the agglomeration coefficient $C$ 
and the average distance $L$, which are defined as
\begin{equation}
	C= \frac{1}{N}\sum_{i=1}^N \frac{2m_i}{k_i(k_i-1)},
\end{equation}
\begin{equation}
	L= \frac{2}{N(N-1)}\sum_{i,j=1}^{N} d_{i,j},
\end{equation}
where $k_i$ is the number of connections of node $i$, 
$m_i$ is the number of connections between its nearest neighbors, 
and $d_{i,j}$ is the shortest distance between nodes $i$ and $j$.

Depending on the value of $p$,  the quantities $C$ and   $L$ change, decaying to zero 
as $p$ increases, as shown in Fig.~\ref{fig:CeL}, for different sizes $N$ and 
average connectivity $\langle k \rangle$.
The SW property,  characterized by high agglomeration and low average distance $L$, 
emerges for intermediate values of $p$, and can be defined as follows

\begin{eqnarray}
\label{eq:p1p2}
	p>p_1(\epsilon) &\longrightarrow & L(p)/L_0 < \epsilon, \nonumber\\
	p<p_2(\epsilon) &\longrightarrow & C(p)/C_0 > (1-\epsilon), 
\end{eqnarray}
where $p_1$ and $p_2$ are the values of $p$ for which $ L(p)= L_0 \epsilon$ 
and $C(p)= C_0 (1-\epsilon)$, respectively, for a given value  $0<\epsilon<1$. 
 Although there is no precise choice for $\epsilon$,   
setting $\epsilon=0.2$, we find the SW regions, which are shadowed
in Fig.~\ref{fig:CeL}.

\section{Opinion  dynamics}
\label{sec:dynamics}

The state of each $i$ node, $1\le i \le N$, is described by  the variable $S_i$, that can take values 
$s_1,\ldots,s_q$, representing the $q$ different opinions. It can also take the 
value $ s_0 $, when the individual does not adopt a defined option. 
 
We start the dynamics with a network where all nodes have not a formed opinion yet, 
that is, $S_i= s_0$ for all nodes $i$. 
Then, we  attribute opinions to randomly chosen nodes called ``initiators''. 
In order to make all alternatives equivalent, we consider the same number of initiators ($I$) 
for each opinion, hence, there is a total of $qI$ initiators.
Thereafter, opinions evolve according to two different rules, whose main difference is the 
direction of the contagion flow: one from one node to the neighbors (TF dynamics) and the other from the 
neighborhood to a central node (PR dynamics), as depicted in Fig.~\ref{fig:flows}.

\subsection{TF dynamics}
\label{sec:TF}

At each Monte Carlo Step (MCs), we apply $ N $ times the following rules:

\begin{itemize}  
	\item Select at random  a  node $i$ (whose opinion is given by the value of $S_i\neq s_0$).
	\item For every neighbor $j$ of node $i$, 
	i) if $S_j=s_0$, then $S_j$ takes the value of $S_i$;
	ii) if $S_j=S_i$, then $S_j$ remains the same;
	iii) if $S_j \neq S_i$ and $S_j \neq s_0$, then $S_j$ assumes the value of 
	$S_i$ with  probability  $r$. 
\end{itemize}

The update of $S_j$ is done after each interaction (asynchronous update).
This kind of dynamics is based on the assumption that undecided individuals are usually passive, 
in the sense that they do not spread their lack of opinion, 
while undecided individuals are easily convinced by interaction with someone who already has a formed opinion.
In addition, the flexibility to change opinion due to an interaction 
is quantified by parameter $r$, which for simplicity adopts the same value for all individuals, 
according to the original version of the model~\cite{travieso2006spread}.

For this dynamics, we measure the quantities of interest in a quasi-stationary state. 
(Actually, in a finite system, the final stationary macrostate, 
after a transient time interval that depends on the model parameters, is the consensus of all nodes.)
We consider the beginning of the quasi-stationary regime as the time when  
the fraction of undecided nodes falls to zero.
Then, we measure the quantities of interest at twice that time.

\subsection{PR dynamics}
\label{sec:PR}

In this dynamics, at each Monte Carlo Step, we visit all the nodes of the network in a random order. 
The status of the network is updated according to the following steps:

\begin{itemize} 
	\item We define the set of nodes ${\cal A}_i $ constituted by $i$ and its nearest neighbors in the network.
	\item We determine the plurality state $s$, 
	as being the state (different from $s_0$) shared by  the largest number  of nodes in this 
	set ${\cal A}_i$ 
	(including  the current state of agent $i$).
	\item Agent $i$ then adopts the plurality state $s$.
  \item In case of a tie, the agent $i$ does not change opinion.
\end{itemize}

In addition, the update can be done in two different modes: asynchronous or synchronous. 
In the first case,  states are updated instantly after each interaction, like in the TF dynamics. 
In the second case, all states of nodes in the network are updated simultaneously after performing $N$ 
interactions (one Monte Carlo step). 
Updates are repeated until the system attains a final state, which is an absorbing state for this dynamics. 

\begin{figure}[!ht]
\begin{center}
	\includegraphics[width=0.5\textwidth]{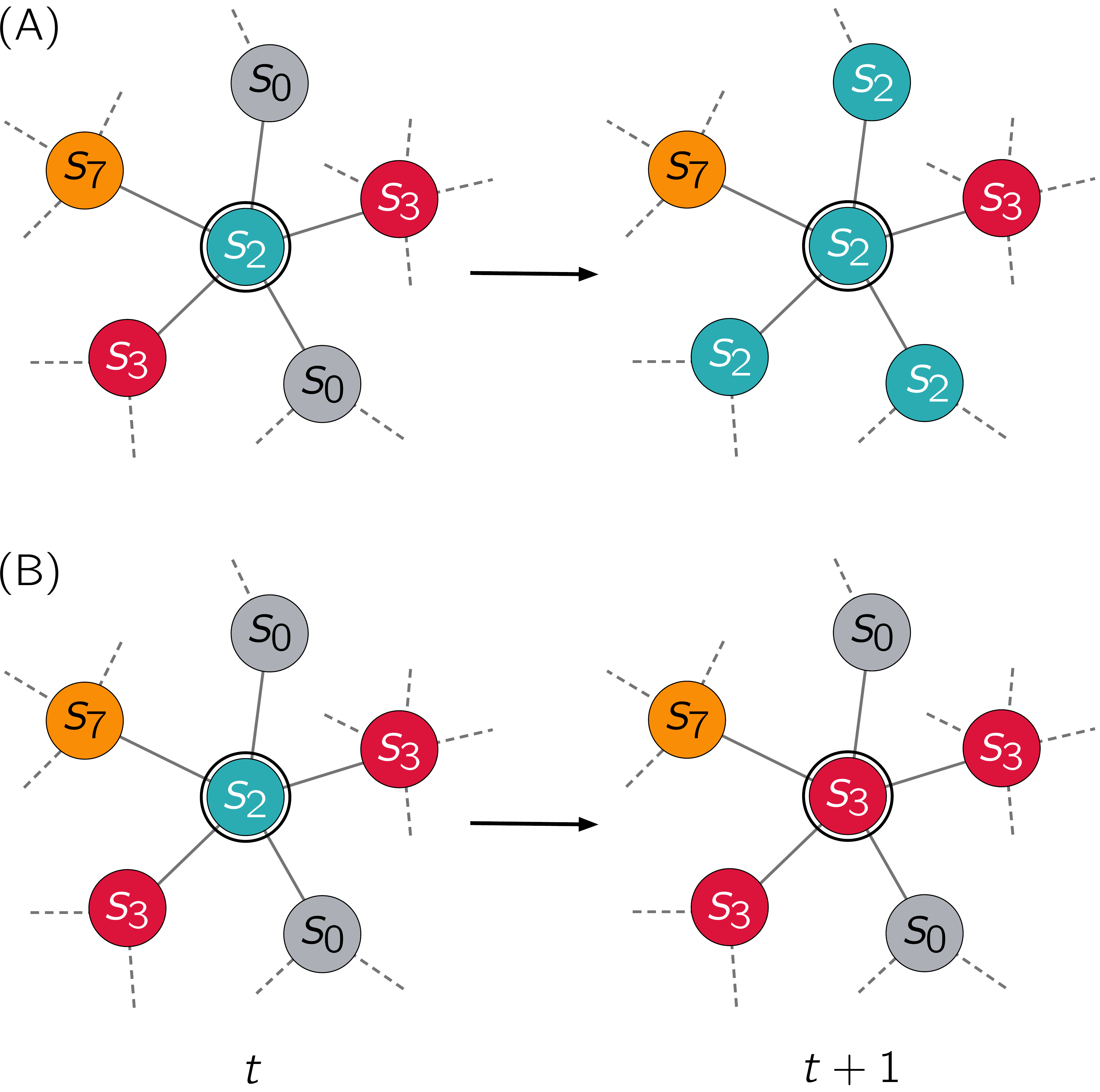}
	\caption{
	(\textbf{A}) TF dynamics: a node $i$ (the central node - blue) spreads its opinion at time $t$ 
	to undecided neighbors and, with a probability $r$, the decided neighbors change their opinion at $t+1$.
	(\textbf{B}) PR dynamics: at a given instant $t$, a node $i$ (the central node - blue) 
	and its neighbors form  a group ${\cal A}_i$, whose plurality state,  
	in the case shown in the figure, is $s_3$, since this is the opinion shared by more nodes. 
	So, in the next step, $t+1$, the $i$ node changes its opinion from $s_2$ to $s_3$.} 
	\label{fig:rules}
\end{center}
\end{figure}

In Fig.~\ref{fig:rules} we depict the contagion dynamics for the two rules used.

\section{Results}
\label{sec:results}

For each realization of the dynamics, we measure the fraction of nodes 
that share the most popular opinion $s_w$, or winning choice,

\begin{equation} 
f_{w}\equiv \frac{n_{s_w}}{N} , 
\end{equation}  
where $n_{s_w}$ is the number of nodes with opinion $S=s_w$.

We also measure the fraction of undecided individuals  as

\begin{equation} 
f_0\equiv  \frac{ n_{s_0}}{N},    
\end{equation} 
where $n_{s_0}$ is the number of nodes with $S=s_0$.

These two quantities give average  information on how opinions are scattered across the population.

\subsection{Time evolution of opinions}
\label{sec:evolution}

\begin{figure}[b!]
	\begin{center}
		\includegraphics[width=0.7\textwidth]{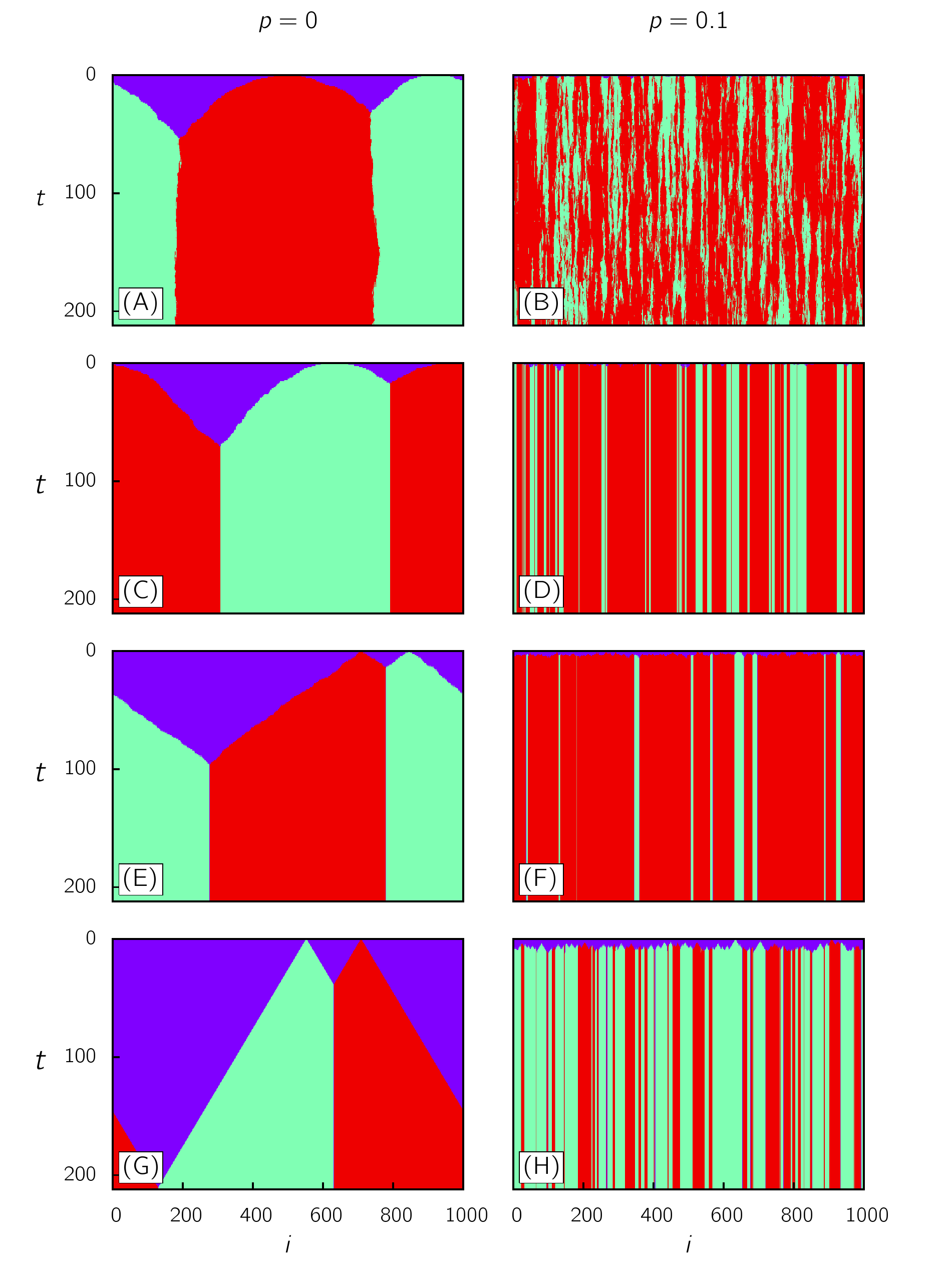}
	\end{center}
	\caption{Graphical representation of the opinion dynamics  
	for networks with $p=0$ (left-hand-side panels) and $p=0.1$ (right-hand-side panels), 
	$N=10^3$, $\langle k \rangle=4$ and $q=2$. 
	Different opinions are represented by different colors, 
	where purple equals $s_0$. 
	In the abscissa axis, we represent the individuals of the population and, 
	in the ordinate axis, the time in MC steps.
	Different dynamic rules are considered, as follows.   
	 {\footnotesize \textsf{(A)}}-{\footnotesize \textsf{(B)}}  TF dynamics for $r=0.1$;
	 {\footnotesize \textsf{(C)}}-{\footnotesize \textsf{(D)}} TF dynamics for $r=0$;
	 {\footnotesize \textsf{(E)}}-{\footnotesize \textsf{(F)}}  asynchronous PR dynamics;   
	 {\footnotesize \textsf{(G)}}-{\footnotesize  \textsf{(H)}} synchronous PR dynamics. 
	}
	\label{fig:evolution}
\end{figure}

A graphical representation of the time evolution  for each rule is provided 
in Fig.~\ref{fig:evolution}, 
where we plotted the state of each node (each opinion is represented by a different color) 
as a function of  time $t$, recalling that each time unit (MCs) 
is accomplished by $N$ iterations of the TF or PR rules. 
In  Fig.~\ref{fig:evolution},   the rules were simulated for $N=10^3$, $\langle k \rangle=4$, 
$q=2$ available options, besides the undecided state. 
One initiator per option ($I=1$) is considered.
This might correspond to the real situation of two different products of 
similar attractiveness being released on the market.
Initially, there is predominance of undecided nodes (purple) 
that diminish with the passage of time.

A  general observation  is that for regular networks ($p=0$, left-hand-side panels), 
the options take longer to propagate than in networks with random connections ($p=0.1$, right-hand-side panels).  
Decreasing the average distance promotes to reach faster the undecided nodes and influence them to take a 
determined opinion.

\begin{figure}[h!]
	\begin{center}
		\includegraphics[width=0.8\textwidth]{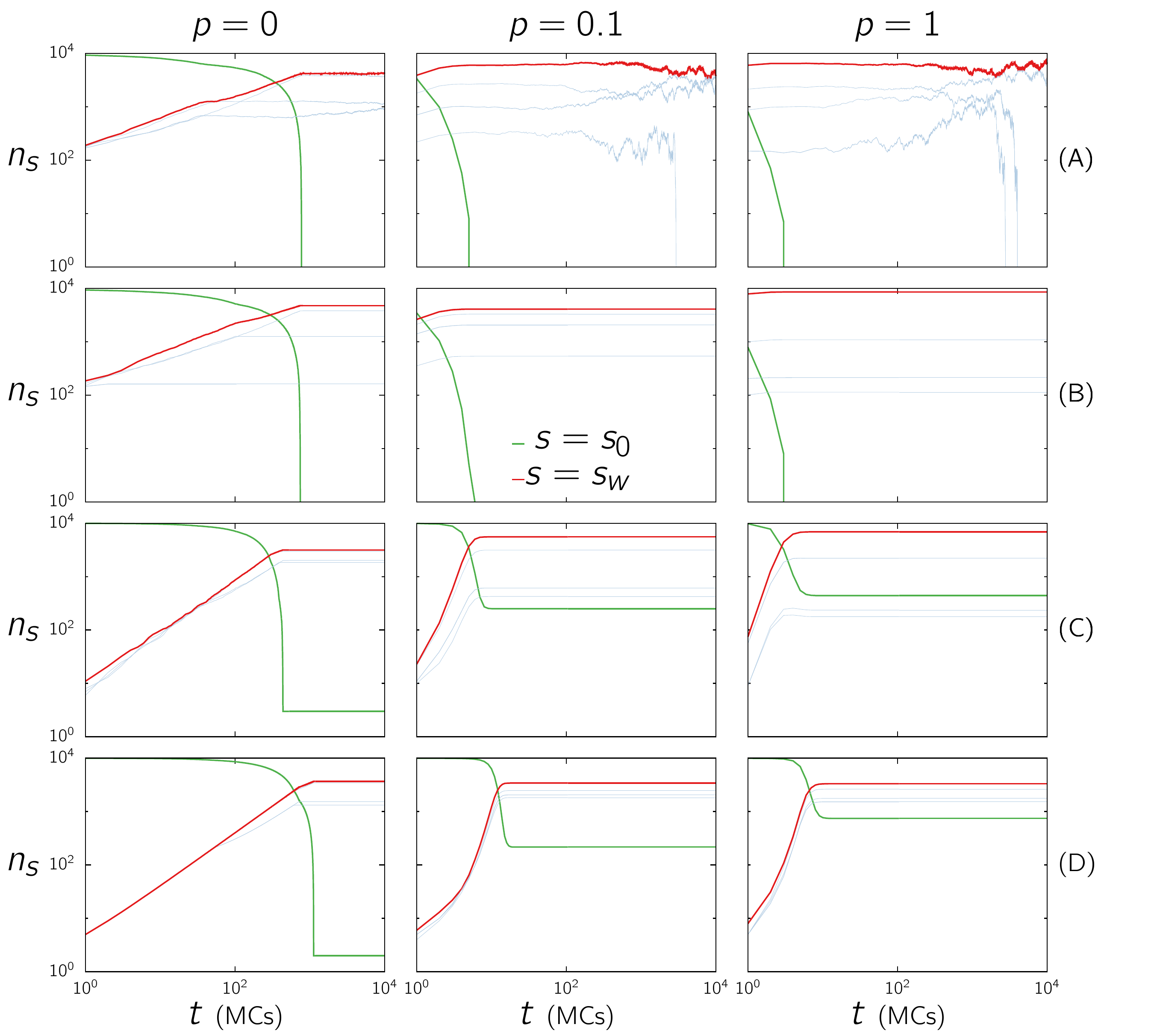} 
	\end{center}
	\caption{Temporal evolution of opinion dynamics for networks with $p=0$ (regular), $p=0.1$ (SW) and $p=1$ (random),  with $N=10^4$, $\langle k\rangle=4$ and $q=4$. 
	The total number of nodes with the winning opinion (red), and of undecided node (green) are shown. 
	{\footnotesize \textsf{(A)}} TF dynamics with  $r=0.1$; 
	{\footnotesize \textsf{(B)}} TF dynamics with $r=0$; 
	{\footnotesize \textsf{(C)}} asynchronous PR dynamics; 
	{\footnotesize \textsf{(D)}} synchronous PR dynamics. }
	\label{fig:nsxt}
\end{figure}

The evolution of the total number of individuals sharing each opinion $s$ 
is shown in Fig.~\ref{fig:nsxt}, 
highlighting those with the winning opinion ($n_{s_w}$, in read) 
and undecided ones ($n_{s_0}$, in green), 
for a network with $N=10^4$ nodes, $\langle k \rangle=4$ and $q=4$.

Panel (A) in Fig.~\ref{fig:nsxt} corresponds  to TF dynamics with $r=0.1$. 
We can observe the existence of fluctuations as in the Fig.~\ref{fig:evolution}. 
A quasi-stationary state is reached when the number of undecided individuals goes to zero. 
The final state is consensus (even if it is not observed within the represented time interval).
Panel (B) corresponds  to TF dynamics with $r=0$. 
In the fully random network we see that one of the possible choices got the majority of followers.
For the TF dynamics the undecided state always disappears, in contrast to the PR dynamics shown in 
panels (C) and (D), with asynchronous and synchronous updates, respectively.

In the following subsections, the fractions $f_w$ and $f_0$ will 
be measured in the quasi-stationary (TF) or final absorbing (PR) macrostates of the system. 
In all cases, unless stated otherwise, the fractions were 
computed averaging over  $10^3$ realizations.

\subsection{Effects of network randomness}
\subsubsection{TF dynamics}

\begin{figure}[!ht]
	\begin{center}
		\includegraphics[width=1\textwidth]{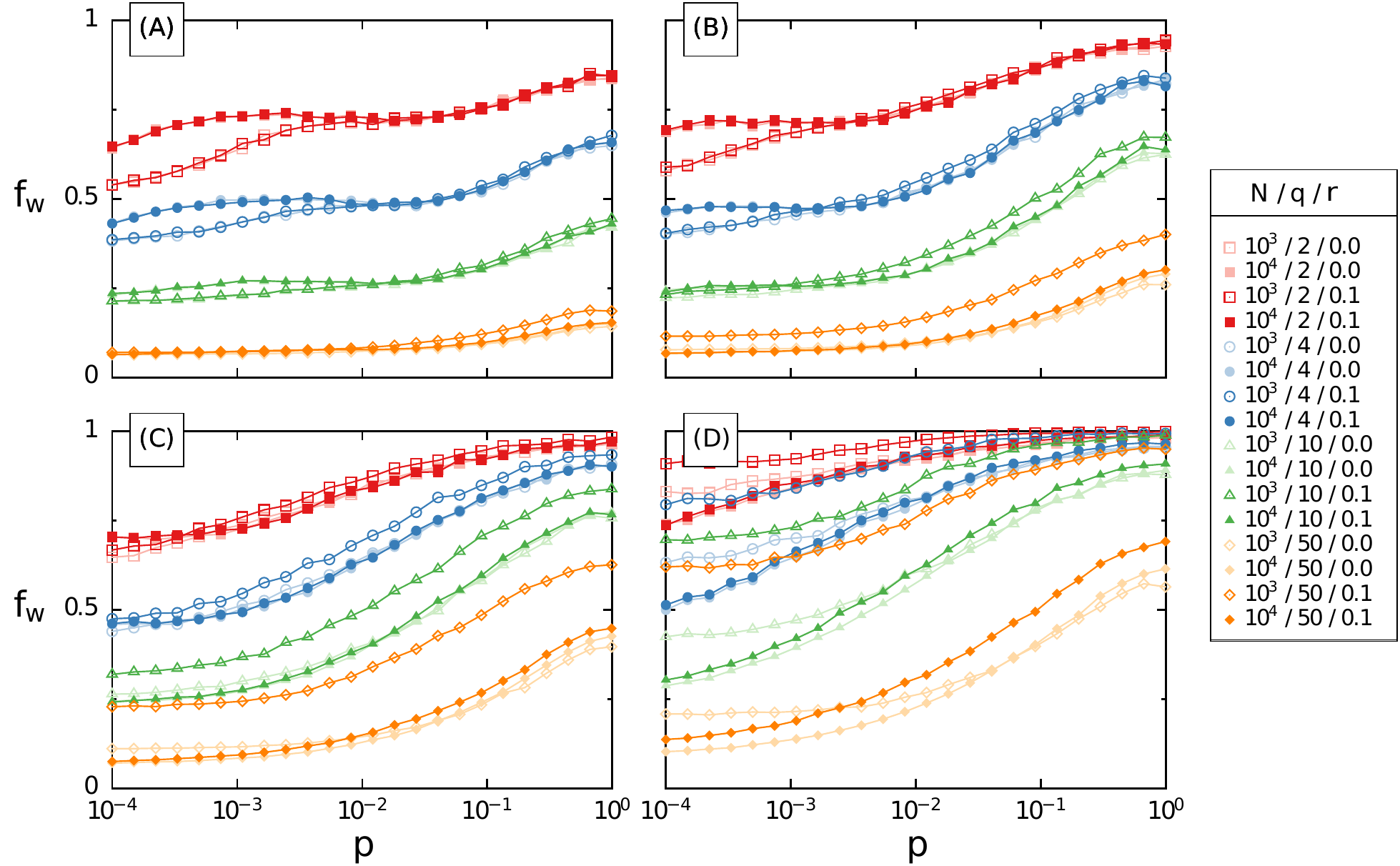}
	\end{center}
	\caption{TF dynamics in WS networks. 
	Average winning fraction $f_w$ as a function of randomness parameter $p$, 
	for  different values of $N$ (hollow or filled symbols), 
	$q$ (different colors) and $r$ (different shades of the same color). 
  Values were averaged over $10^3$ realizations. 
	{\footnotesize \textsf{(A)}} $\langle k\rangle=4$; 
	{\footnotesize \textsf{(B)}} $\langle k\rangle=10$; 
	{\footnotesize \textsf{(C)}} $\langle k\rangle=20$; 
	{\footnotesize \textsf{(D)}} $\langle k\rangle=50$. 
	}
	\label{fig:TF}
\end{figure}

The effects of  randomness parameter $p$ on the fraction $f_w$,  for the TF rule, 
are shown in Fig.~\ref{fig:TF}. 
Panels (A)-(D) correspond  to different values  of the average connectivity $\langle k\rangle$. 
Moreover, in each panel, results are shown for two different sizes $N$: 
$10^3$ (hollow symbols) and $10^4$ (filled symbols). 

As expected, the winning fraction $f_w$ decreases as the number of option $q$ increases. 
 We observe that, when there are more reconnections in the network (larger $p$), 
the value of $f_w$ becomes higher.
Moreover, consensus becomes more likely when the network connectivity  $\langle k \rangle$ 
increases. 
This can be understood as follows. 
In a regular (small $p$) and  low connected (small  $\langle k \rangle$) 
network,  most initiators have a similar connectivity, then opinions 
spread more homogeneously, hence $f_w \approx 1/q$. 
As $p$ and/or   $\langle k \rangle$ increase,  the dispersion of connectivities increases, 
therefore,  a  highly  connected initiator  will have more chance to dominate. 
As a consequence, $f_w$ grows and consensus becomes more likely.

\subsubsection{PR dynamics}

\begin{figure}[!h]
	\begin{center}
		\includegraphics[width=1\textwidth]{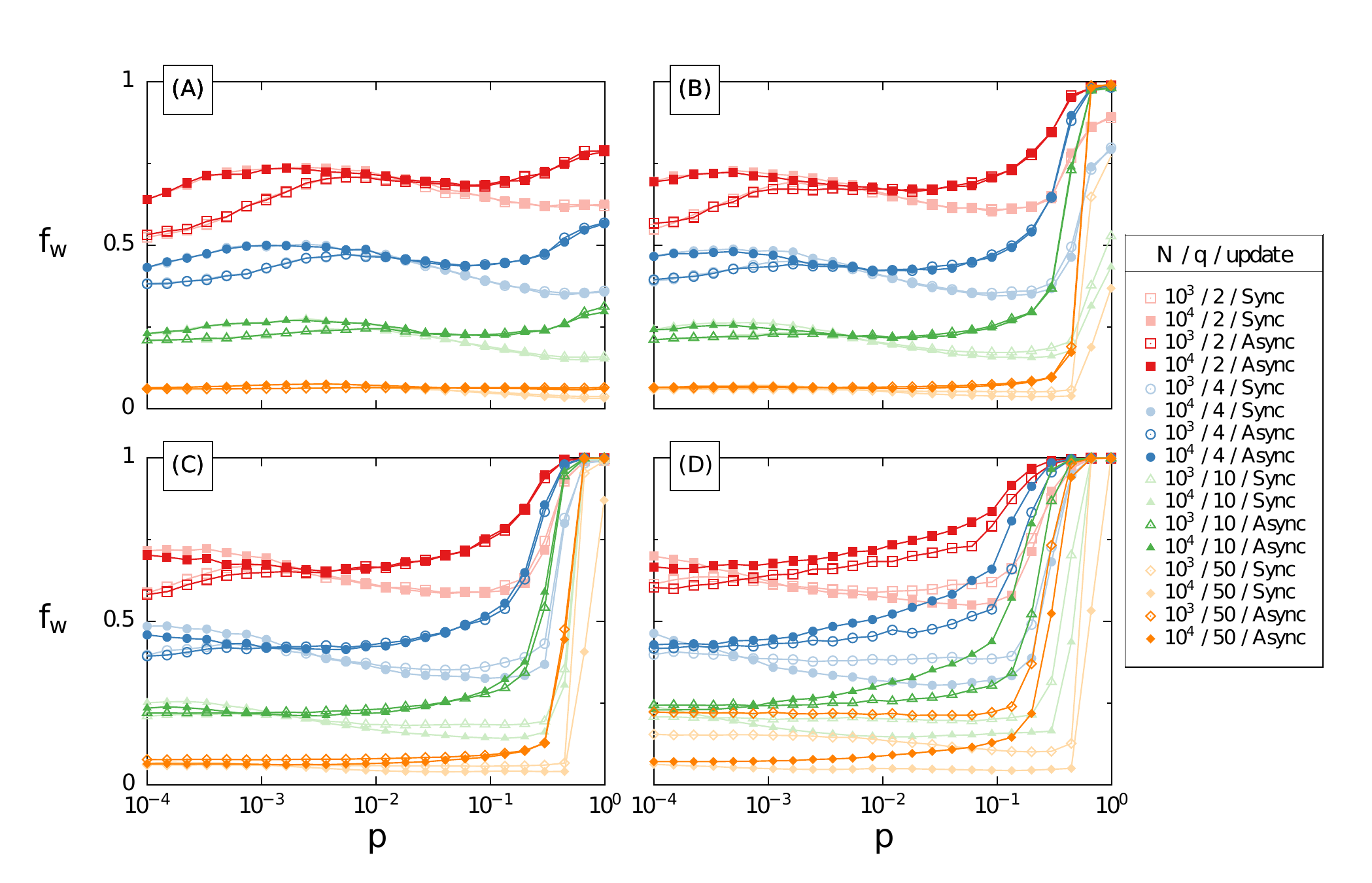}
	\end{center}
	\caption{PR dynamics in  WS networks. 
	Average winning fraction $f_w$ as a function of the randomness parameter $p$,  
	for different values of $N$ (hollow and filled symbols), $q$ (different colors) 
	and two updates (synchronous and asynchronous).  
 {\footnotesize \textsf{(A)}} $\langle k\rangle=4$; 
{\footnotesize \textsf{(B)}} $\langle k\rangle=10$; 
{\footnotesize \textsf{(C)}} $\langle k\rangle=20$; 
{\footnotesize \textsf{(D)}} $\langle k\rangle=50$.
}
	\label{fig:PR-fmax}
\end{figure}

The behavior of $f_w$ versus $p$ for this dynamics is presented in Fig.~\ref{fig:PR-fmax}.  
Panels (A)-(D) correspond  to different values  of the average connectivity $\langle k\rangle$. 
Moreover, in each panel, results are shown for two different sizes $N$: 
$10^3$ (hollow symbols) and $10^4$ (filled symbols). 
In this case, the outcomes for two kinds of updates (synchronous and asynchronous) are shown.
 
At very small $p$,  the effect  of network connectivity  is almost negligible.
Moreover, up to  $p \simeq  0.1$,  $f_w$ is weakly dependent  on $p$, 
remaining at  a value slightly above $1/q$, indicating equipartition of opinions.
This property is distorted for small  number of options. 

For small values of $p$, we observe that $f_w$ is very close for both updates. 
However, by increasing the value of $p$, the asynchronous update facilitates 
the predominance of one of the options, 
and in some cases promotes consensus, even for small connectivity. 
Note the abrupt increase of $f_w$ occurring above $p=0.1$ in panels (B)-(D).

\begin{figure}[!ht]
	\begin{center}
		\includegraphics[width=1\textwidth]{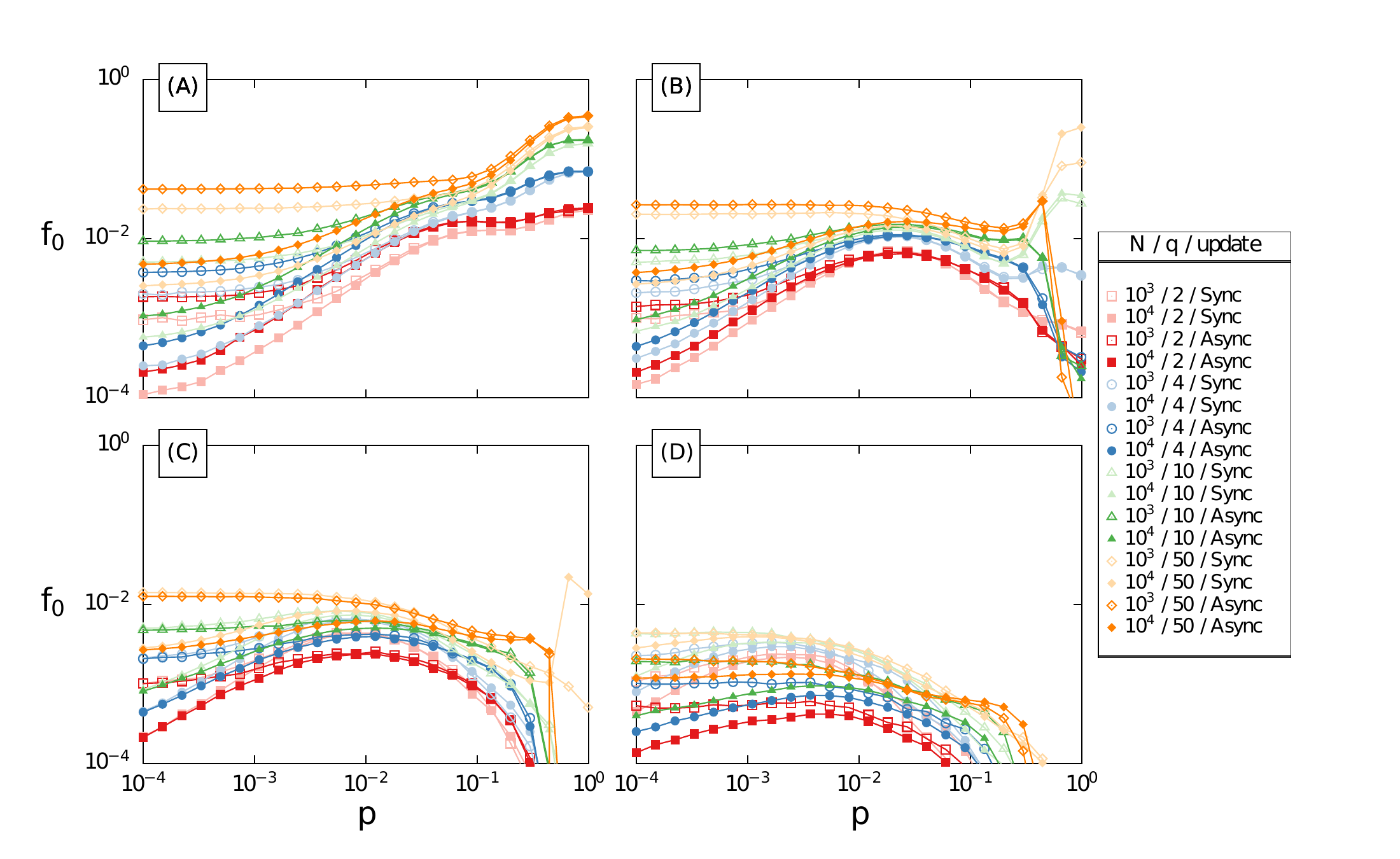}
	\end{center}
	\caption{PR dynamics in  WS networks. 
	Average undecided fraction $f_0$ as a function of the randomness parameter $p$. 
		All settings are the same of Fig.~\ref{fig:PR-fmax}. 
		{\footnotesize \textsf{(A)}} $\langle k\rangle=4$; 
{\footnotesize \textsf{(B)}} $\langle k\rangle=10$; 
{\footnotesize \textsf{(C)}} $\langle k\rangle=20$; 
{\footnotesize \textsf{(D)}} $\langle k\rangle=50$.
		}
	\label{fig:PR-f0}
\end{figure}

For this dynamics, we also show the fraction of undecided nodes $f_0$ as a function of $p$, 
in  Fig.~\ref{fig:PR-f0}.

For sufficiently large connectivity $\langle k \rangle$  and  small number of options $q$ 
 [panels (B)-(D)], 
there is  a maximum value of the undecided fraction  in the SW region. 
This region is below $p\simeq 0.85$, 
almost independently of $N$ and $\langle k \rangle$, and 
its lower bound decreases with $\langle k \rangle$, as illustrated 
in Fig.~{\ref{fig:CeL}}.

But as $\langle k \rangle$ grows, the undecided nodes become a very small fraction of the population. 
As commented earlier, 
the dispersion of the mean   connectivity increases with $\langle k \rangle$, 
and for this reason,  the occurrence of a tie in the opinions of the neighbors of a given node 
is more difficult. This effect is enhanced by increasing randomness, and  $f_0$ becomes negligible for $p>0.1$ (outside the SW region).
The presence of the local maximum is associated with the value of $p$ where there 
is more ``competition'' at this stage of the evolution of opinions. 

Also note that  Fig.~\ref{fig:evolution} shows that undecided nodes  
are present right at the interfaces of different opinions.

Smaller networks make it easier  indecision to occur, and the more options are available, 
more undecided nodes  are present on the network, because ties are more frequent, 
typically many opinions with one representative.

\subsection{Dependence with the connectivity}

\begin{figure}[!ht]
	\begin{center}
		\includegraphics[width=0.9\textwidth]{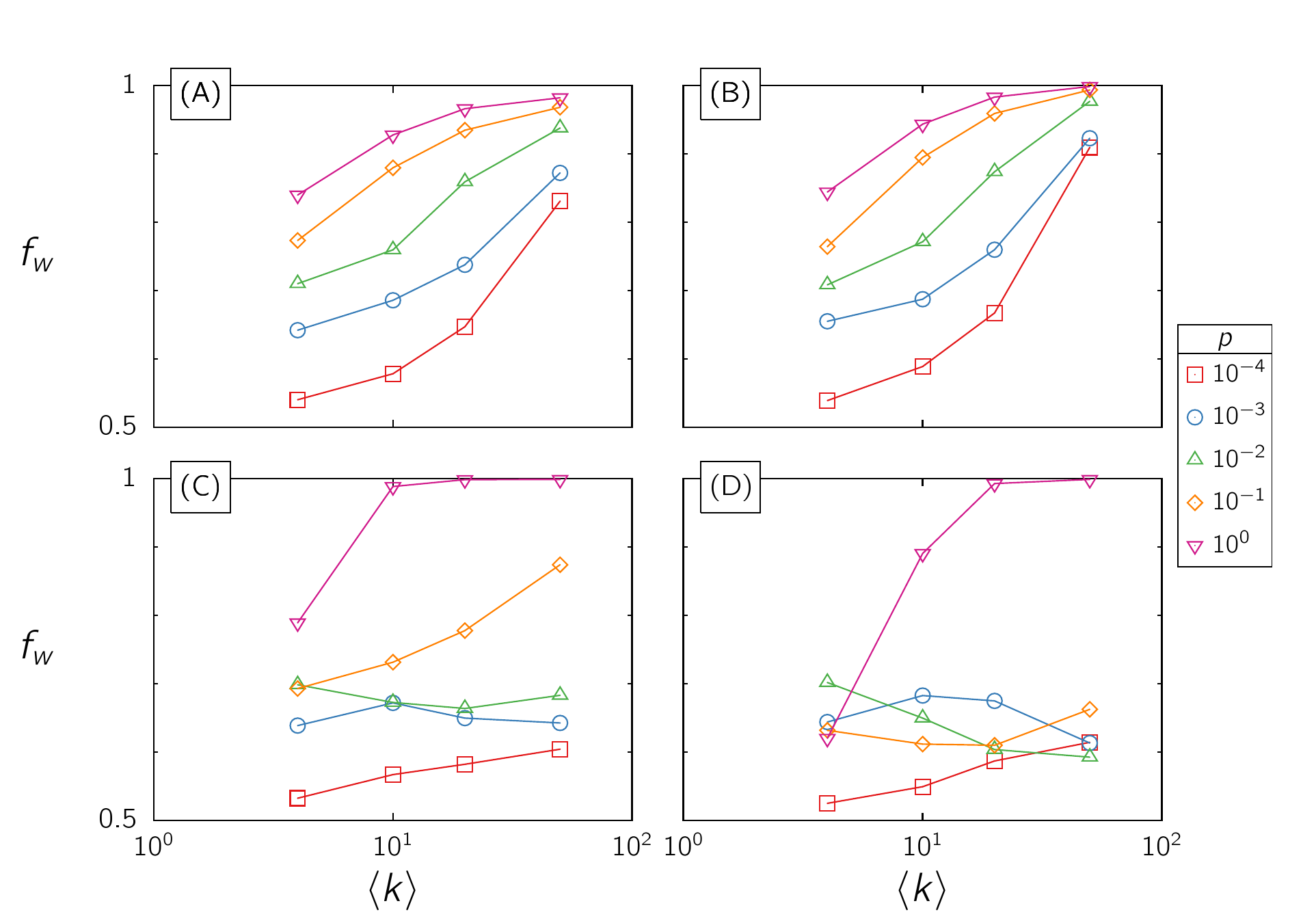}
	\end{center}
	\caption{ Average winning fraction $f_w$ versus the network connectivity, 
	 for $N=10^3$ and $q=2$; 
	{\footnotesize \textsf{(A)}} TF dynamics, with $r=0$;
	{\footnotesize \textsf{(B)}} TF dynamics, with $r=0.1$;
	{\footnotesize \textsf{(C)}} asynchronous PR dynamics and 
	{\footnotesize \textsf{(D)}} asynchronous PR dynamics. 
	Each color corresponds to a different value of $p$.}
	\label{fig:fmax_k}
\end{figure}

\begin{figure}[!ht]
	\begin{center}
		\includegraphics[width=0.9\textwidth]{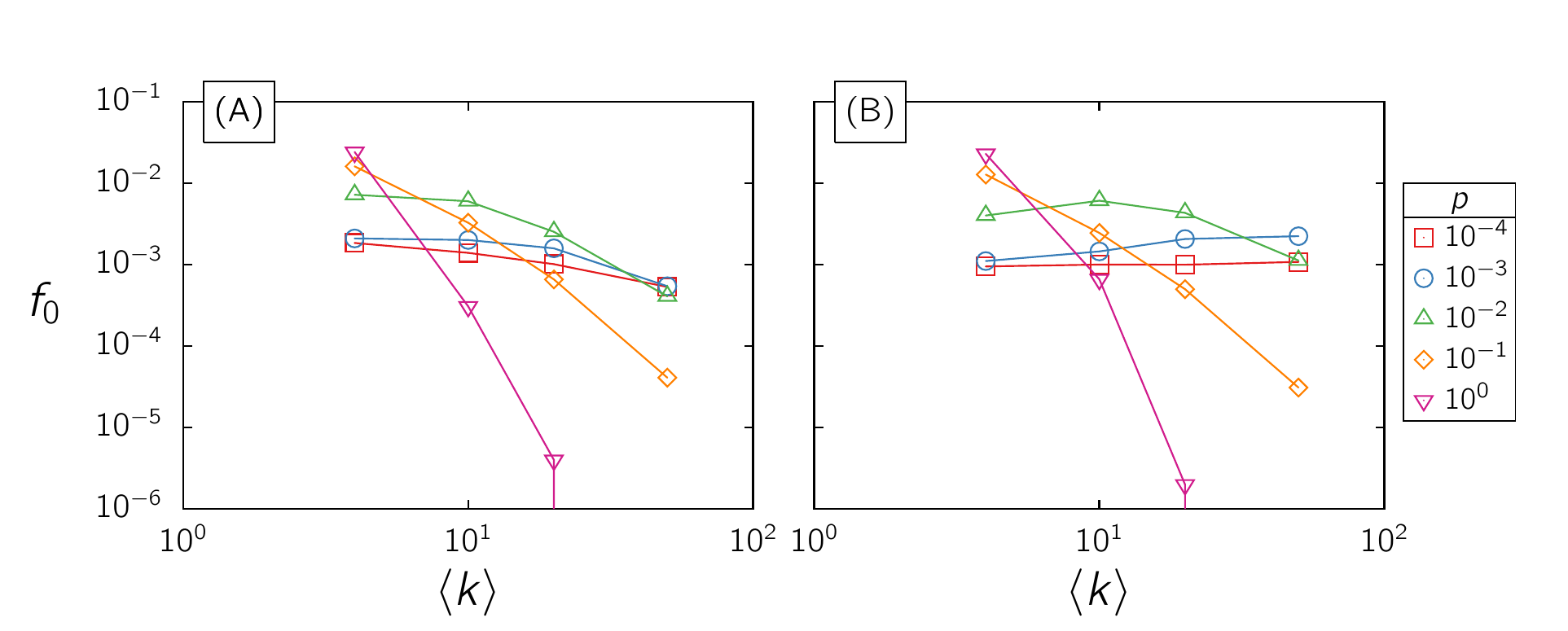}
	\end{center}
	\caption{Average undecided fraction $f_0$ versus the network connectivity, 
	 for $N=10^3$ and $q=2$; using PR dynamics with 
	{\footnotesize \textsf{(A)}} asynchronous and {\footnotesize \textsf{(B)}} synchronous updating. 
	Each color corresponds to a different value of $p$.}
	\label{fig:f0_k}
\end{figure}

In order to better understand how network properties affect the final settings of state variables, 
we represent in the Figs.~\ref{fig:fmax_k} and \ref{fig:f0_k} 
the fractions of the winning opinion and the fractions of undecided ones as a function of the network average connectivity, 
for different rules and values of $p$.

For the TF dynamics, images {\footnotesize \textsf{(A)}} and {\footnotesize \textsf{(B)}} of  Fig.~\ref{fig:fmax_k}, it is notorious the growth of the winning fraction as a function of $\langle k \rangle$, and also, as seen previously, that a larger fraction of supporters of the winning opinion 
is directly related to higher randomness.

For the PR dynamics, images {\footnotesize \textsf{(C)}} and {\footnotesize \textsf{(D)}}, the curves corresponding to $p$ equal to $10^{-1}$, $10^{-2}$ and $10^{-3}$,  which are all in the SW region, where non-trivial behavior occurs.

In Fig.~\ref{fig:f0_k} we represent the fraction of undecided nodes $f_0$
as a function of the average network connectivity for the asynchronous 
and synchronous updates, respectively, using the dynamics PR. 
For small $ p $, $f_w$ is almost unaffected by the network connectivity, 
due to the existence of a non-competitive regime, 
where nodes only interact with undecided nodes. 
Something similar was observed in regular lattices~\cite{calvao2016role}.
For larger $p$, the number of undecided nodes decreases with $\langle k \rangle$ 
because, as we commented before, due to large connectivities and network dispersion, 
it would be more difficult for an individual to be in a situation where neighbors' nodes tie.

\subsection{Influence of the number of initiators}

\begin{figure}[!ht]
	\begin{center}
		\includegraphics[width=0.9\textwidth]{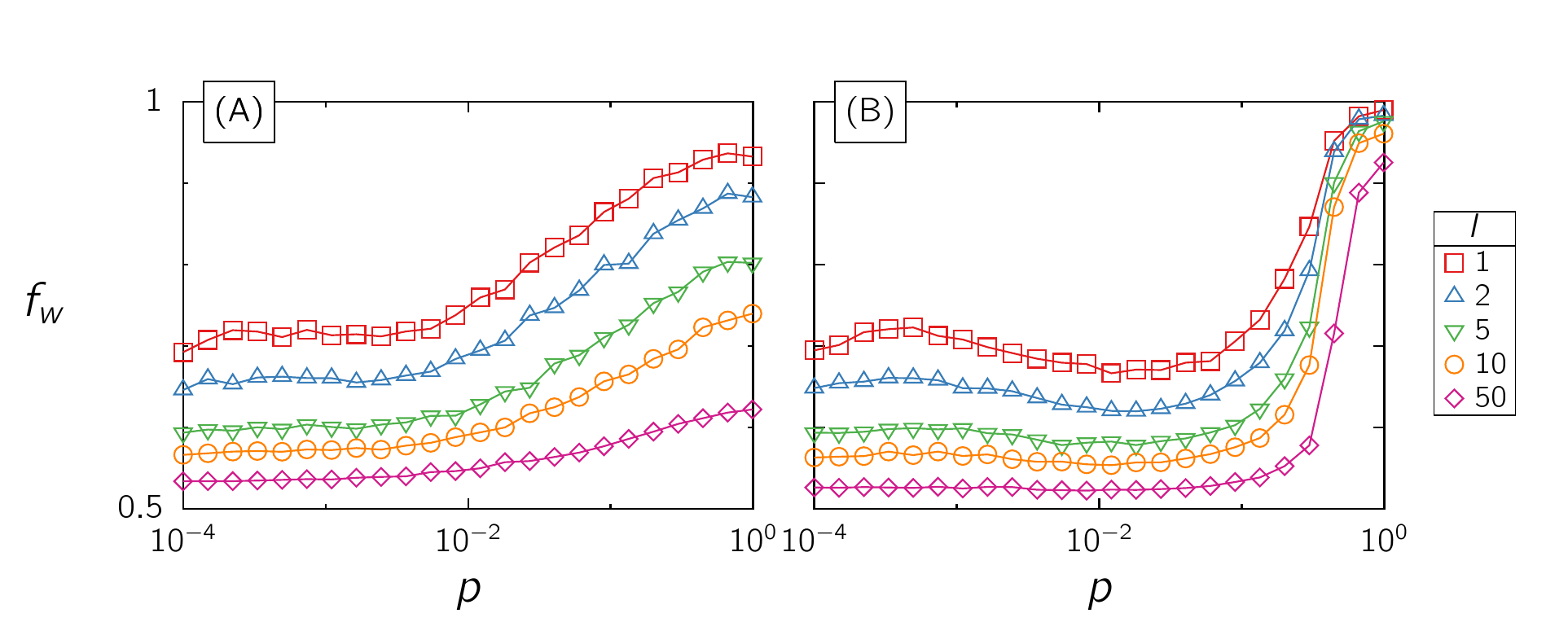} 
	\end{center}
	\caption{Average winning fraction $f_w$ versus $p$, 
	for different number  of initiators per opinion, 
	using $N=10^4$, $\langle k \rangle=10$ and $q=2$;
	{\footnotesize \textsf{(A)}} TF dynamics with $r=0.1$ and {\footnotesize \textsf{(B)}} PR dynamics with asynchronous update.}
	\label{fig:differentI}
\end{figure}

In all the results shown so far, it was assumed that each option was introduced into the network 
by only one individual ($I=1$). 
But what happens in the presence of more initiators? 
In Fig.~\ref{fig:differentI}, we show the results for 
TF dynamics with $r=0.1$ and  PR dynamics with asynchronous update.

As we increase the amount of opinion propagators in the initial configuration, 
we observe a decrease of $f_w$ for all $p$.

\section{Final remarks}
\label{sec:conclusions}

We considered two simple rules (TF and PR) that capture essential features of a multi-state opinion dynamics, 
mimicking the scenario where individuals have to choose one between several options.  
Nodes, representing individuals, are on top of a SW network, with a given average 
connectivity $\langle k \rangle$ and a proportion $p$ of random shortcuts.

Opinion dynamics (for each rule) starts with the majority of nodes in the undecided state, 
while a few nodes (initiators) have a defined opinion. 
Initiator nodes are chosen at random, irrespective of their connectivity, 
or of any other centrality measure.

Our main interest was to understand how properties of the network, 
such as average connectivity and  proportion of random shortcuts, 
affect the final distribution of opinions.

Following TF dynamics, all nodes become decided at the final state, but we analyzed quase-steady states. 
Stronger randomness, as well as higher average connectivity, increases the chance of a predominant opinion. 

Meanwhile, following  PR dynamics, there remain undecided nodes in the stationary state.
These undecided nodes reside on the interface of clusters of nodes with the same opinion. 
For small $p$, each cluster first grows independently (non competitive phase), then enters 
a competitive regime up to the time when the dynamics becomes frozen due to ties.
Increasing the number of shortcuts on the network also enhances these interfaces (see Fig. 4), 
therefore, there is increase of the fraction of undecided nodes with $p$. 
One important effect to be considered is the dispersion of node connectivities. 
A large dispersion makes hard to have ties. Once the dispersion increases, 
the dynamics enters in a competitive regime, and neighbouring clusters invade one another. 
This could also result in a predominant opinion, that can dominate the whole network, giving rise to consensus.

On conclusion, we see that when decisions are made through a group of people, as in PR dynamics, 
there is more chance that indecision prevails. 
Differently, when it is an individual that influences its neighbours, as in TR dynamics, 
undecided nodes are absent in the final state.

As a perspective of future work, it will be interesting to know how the results change 
if initiators are pick-up by some predefined criterion, like preferential spreader sites~\cite{Kitsak2010}.


\bibliographystyle{elsarticle-harv} 

\end{document}